\documentclass[aip,reprint,amsmath,amssymb,pof,onecolumn,10pt]{revtex4-2}
\usepackage{graphicx}% Include figure files
\usepackage{dcolumn}% Align table columns on decimal point
\usepackage{geometry}
\usepackage{hyperref}
\hypersetup{
    colorlinks=true,
    linkcolor=blue,
    citecolor=blue,
    filecolor=magenta,
    urlcolor=blue
}
\usepackage{multirow}
\usepackage{mathtools}
\usepackage{float}
\usepackage{xcolor}
\usepackage{booktabs}
\usepackage[dvipsnames]{xcolor}
 
\usepackage{xr}
\usepackage{natbib}
\usepackage{lipsum}
\usepackage{tikz}
\usetikzlibrary{arrows.meta, positioning, shapes.geometric}
\begin{document}
	
	\title{Data-Driven Surrogate Modeling for Micromixing of Non-Newtonian Fluids in Sinusoidal Converging–Diverging Microchannels}
	\author{Kritagya Sharma}
	\affiliation{Department of Mechanical Engineering, Indian Institute of Petroleum and Energy, Visakhapatnam, Andhra Pradesh, India}

    \author{Bimalendu Mahapatra}
    \email{bimalendum@iisc.ac.in}
	\affiliation{Department of Mechanical Engineering, Indian Institute of Science, Bangalore-560012, India}

	\begin{abstract} 
    Micromixing of non-Newtonian fluids remains challenging because laminar flow at microscales restricts transverse transport primarily to molecular diffusion. In this study, we investigate the transport mechanisms governing passive micromixing of a Carreau--Yasuda fluid in two-dimensional sinusoidal converging--diverging microchannels and develop a surrogate-assisted framework for their multi-objective design. We perform high-fidelity finite-volume simulations by systematically varying the wall-amplitude ratio, phase offset, and wave count under creeping-flow conditions. The results show that successive contraction--expansion units enhance mixing through the combined effects of interface stretching, elevated shear rates, and shear-thinning-induced viscosity reduction. These mechanisms improve scalar transport but simultaneously increase pressure drop, creating an inherent trade-off between mixing performance and hydraulic resistance. To efficiently explore the multidimensional design space, we construct surrogate models from high-fidelity numerical simulations and identify Gaussian Process Regression (GPR) as the most accurate predictor of both the mixing index and pressure drop. Coupling the validated GPR surrogate with Non-dominated Sorting Genetic Algorithm II (NSGA-II) accurately reproduces the Pareto front obtained from the high-fidelity simulations and identifies optimal microchannel geometries that balance mixing enhancement against pressure loss. The proposed machine learning framework provides a fast, accurate, and physically consistent strategy for the multi-objective design of passive micromixers for non-Newtonian fluids.
    \end{abstract}

	\maketitle

\section{Introduction}

Microfluidic technologies have become indispensable in chemical synthesis~\cite{Elvira2013}, biomedical diagnostics~\cite{Yager2006}, pharmaceutical processing~\cite{DittrichManz2006}, lab-on-a-chip systems~\cite{Mark2010,10.1063/5.0033088,10.1063/5.0003457}, and biomanufacturing~\cite{Headen2014} because they enable precise manipulation of minute fluid volumes with low reagent consumption, rapid heat and mass transfer, and excellent process controllability~\cite{Dittrich2006, Stone2004, Whitesides2006}. Despite these advantages, transport in microchannels is fundamentally constrained by the low-Reynolds-number regime, where viscous forces dominate inertia and the flow remains laminar~\cite{asmolov2018inertial,MAHAPATRA2021104479,mahapatra2022effect,mahapatra2023alterations}. Consequently, transverse transport is governed primarily by molecular diffusion, which is intrinsically slow for liquids and often inadequate to achieve rapid homogenization over practical channel lengths~\cite{Stroock2002,braff2015inertial}. Efficient micromixing therefore remains one of the central challenges in microfluidic engineering, motivating the development of passive geometries capable of enhancing scalar transport without external energy input~\cite{Cai2017, ijms12053263, Soltani2024}.

Efficient mixing under creeping-flow conditions depends on the generation of concentration gradients through the kinematic deformation of the concentration field~\cite{Balasuriya2005}. Stretching and folding of fluid interfaces increase the interfacial area and reduce the characteristic diffusion length, thereby accelerating molecular transport even though the molecular diffusivity remains unchanged~\cite{Schlicketal2013,StoneStone2005}. These mechanisms, commonly described within the framework of chaotic advection, form the physical basis of passive micromixers~\cite{Parketal2024,Yuanetal2022}. Unlike active micromixers, which employ electric~\cite{Mahapatra2022,Mahapatra2024}, magnetic~\cite{10.1039/c2lc40818j}, thermal~\cite{10.1063/5.0178396}, or acoustic forcing~\cite{JANG2007179}, passive devices manipulate the flow solely through geometric modifications of the channel, making them attractive because of their structural simplicity, low energy consumption, and compatibility with conventional microfabrication technologies~\cite{Hanetal2024,GhanbariRahimi2023,MajumdarDasgupta2024,Mahapatra2023}.

Among passive micromixer designs, contraction–expansion microchannels have attracted considerable attention because their periodically varying geometry continuously modifies the local flow deformation, thereby promoting mixing~\cite{10.1063/5.0058732,ijms12053263}. Fluid acceleration through contractions generates regions of high shear rate, whereas subsequent expansions redistribute the velocity field and reorient scalar interfaces before the next contraction~\cite{10.1039/b418314b}. Repeated acceleration–deceleration cycles therefore promote interface stretching, scalar-gradient generation, and diffusion-driven homogenization~\cite{10.1039/b807107a,Mahapatra2024}. Sinusoidal converging–diverging microchannels provide a particularly attractive configuration because the smooth variation of the channel cross-section produces continuous deformation fields while avoiding the abrupt changes associated with sharp contractions and expansions~\cite{10.1063/5.0090190,gepner2023flow}. The wall amplitude, wavelength, and relative phase collectively determine the distribution of shear rate, residence time, and scalar gradients, thereby governing the overall mixing performance~\cite{10.1039/b418314b}. Similar transport mechanisms are encountered in vascular constrictions\cite{10.1063/5.0175142}, porous biological tissues~\cite{10.1063/5.0232176}, extrusion nozzles, and additive-manufacturing flow passages~\cite{10.1063/5.0248530}, where gradual geometric variations strongly influence local hydrodynamics.

Fluids commonly used in microfluidic applications, including blood, polymer solutions, hydrogels, protein suspensions, and bioinks, exhibit pronounced shear-thinning behavior~\cite{Bird1987, Barnes1989,del2017edge}. Unlike Newtonian fluids, their apparent viscosity decreases with increasing shear rate, introducing a strong coupling between the flow kinematics and the local rheology~\cite{HerradaManchon2023,10.1039/d5lc00864f}. In converging--diverging channels, the elevated shear rates generated within contractions reduce the local viscosity, thereby modifying the velocity distribution, residence time, and scalar transport~\cite{Zografos2016}. Consequently, the same geometric features responsible for enhancing mixing also influence the hydraulic resistance, producing an intrinsic trade-off between mixing performance and pressure loss~\cite{Hossain2011, Kaid2024}. The Carreau--Yasuda model has become one of the most widely adopted constitutive descriptions for such fluids because it accurately captures the transition from the zero-shear Newtonian plateau to the high-shear power-law regime while remaining computationally efficient for engineering simulations~\cite{Carreau1972,Yasuda1981,PhysRevFluids.5.044203}. Nevertheless, despite extensive studies on Newtonian passive micromixers, the transport physics of shear-thinning fluids in smoothly varying sinusoidal contraction--expansion geometries remains comparatively unexplored~\cite{10.1063/5.0058732}.

The computational design of passive micromixers remains challenging because high-fidelity computational fluid dynamics simulations, while accurately capturing the coupled transport of momentum and species, require substantial computational resources to optimize multiple geometric design parameters. Recent advances in surrogate modeling provide an efficient alternative by replacing expensive CFD evaluations with data-driven models trained on a limited number of simulations~\cite{Forrester2008, Rasmussen2006, Brunton2020}. Among the available approaches, Gaussian Process Regression is particularly attractive for deterministic CFD datasets because it accurately interpolates smooth nonlinear response surfaces while providing predictive uncertainty~\cite{Shahriari2016}. Although surrogate-assisted optimization has received increasing attention in fluid mechanics, its application to passive micromixers handling non-Newtonian fluids remains limited, particularly in studies that integrate high-fidelity CFD, Carreau--Yasuda fluid transport, and multi-objective optimization~\cite{Brunton2020}.

In this work, we develop a surrogate-assisted framework for the design and optimization of passive micromixers operating with shear-thinning fluids. We investigate incompressible Carreau--Yasuda fluid flow in two-dimensional sinusoidal converging--diverging microchannels and systematically assess the effects of the wall-amplitude ratio, phase offset, and wave number on mixing performance and pressure drop. A database of high-fidelity CFD simulations is used to train and evaluate multiple surrogate models, and the best-performing model is subsequently coupled with the Non-dominated Sorting Genetic Algorithm II (NSGA-II)~\cite{Deb2002} to identify Pareto-optimal microchannel geometries. The optimized designs are subsequently validated using independent CFD simulations, which confirm the surrogate predictions and provide insight into the flow and transport mechanisms governing mixing enhancement. By integrating high-fidelity simulations, surrogate modelling, and multi-objective optimization, this study links geometric modulation to shear-rate amplification, viscosity redistribution, mixing, and hydraulic losses, providing an efficient framework for the design of passive micromixers for complex fluids.

\section{Problem Formulation:}
\begin{figure}[t]
    \centering
    \includegraphics[width=0.95\linewidth]{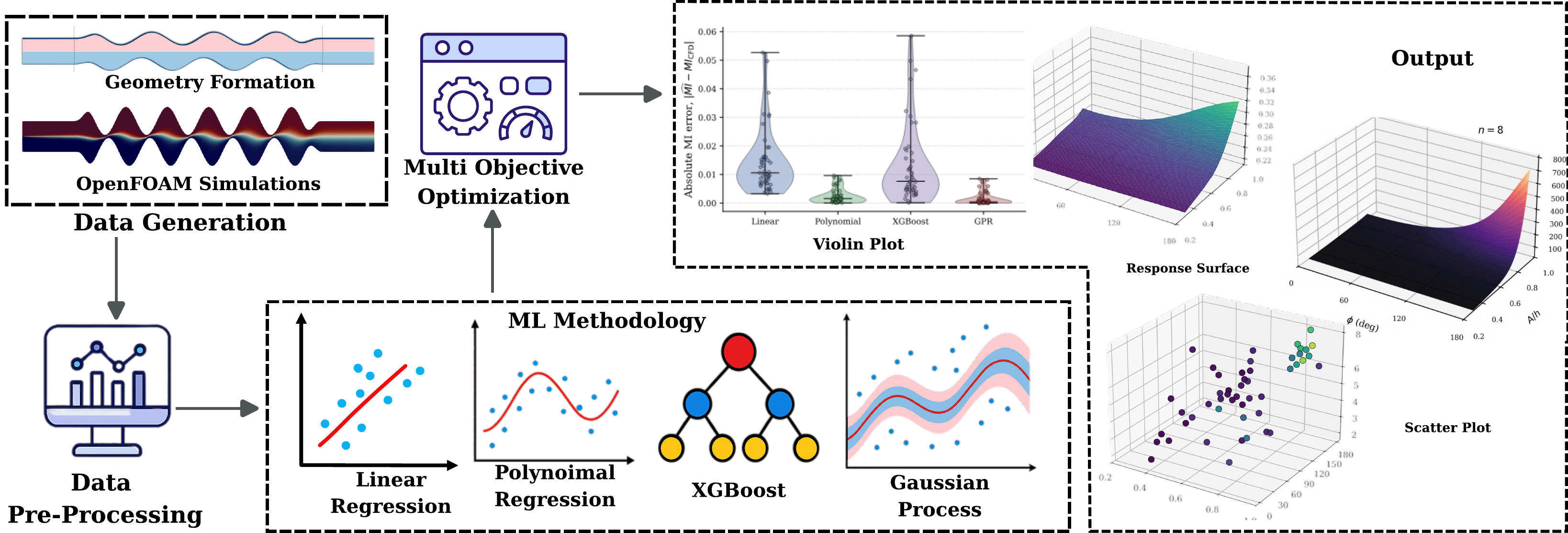}
    \caption{Workflow of the surrogate-assisted optimization methodology, including CFD data generation, machine-learning surrogate development, multi-objective optimization, and model interpretation for Carreau--Yasuda fluid micromixers.}   
    \label{fig:schematic}
\end{figure}

In this study, we investigate the transport and mixing of an incompressible shear-thinning fluid in sinusoidal converging--diverging microchannels. The overall methodology is summarized in Fig.~\ref{fig:schematic}. We combine high-fidelity computational fluid dynamics simulations with surrogate modelling and multi-objective optimization to evaluate the influence of geometric modulation on micromixer performance. Specifically, we quantify the mixing performance and the corresponding pressure drop over a broad range of channel geometries and identify designs that achieve an effective balance between these competing objectives. The following sections describe the physical model, channel geometry, governing equations, numerical methodology, and performance metrics.

\subsection{Geometry Definition}

We consider a two-dimensional sinusoidal passive micromixer consisting of a
straight inlet section, a sinusoidal active section, and a straight outlet
section. The total channel length is $3400\,\mu\mathrm{m}$, comprising a
$500\,\mu\mathrm{m}$ straight inlet section, a
$2400\,\mu\mathrm{m}$ sinusoidal active section, and a
$500\,\mu\mathrm{m}$ straight outlet section. The nominal channel half-height is
$h=100\,\mu\mathrm{m}$. To define the sinusoidal section, we introduce the local streamwise coordinate
$\xi=x-L_{\mathrm{in}}$, where $L_{\mathrm{in}}=500\,\mu\mathrm{m}$ is the inlet
length. Consequently, $0\le\xi\le L_{\mathrm{sin}}$, where
$L_{\mathrm{sin}}=2400\,\mu\mathrm{m}$. The upper and lower channel walls are defined as

\begin{equation}
y_{\mathrm{top}}(\xi)
=
h+A\,E(\xi)\sin\!\left(\frac{2\pi\xi}{\lambda}\right),
\qquad
y_{\mathrm{bot}}(\xi)
=
-h+A\,E(\xi)\sin\!\left(\frac{2\pi\xi}{\lambda}+\phi\right),
\label{eq:walls}
\end{equation}

where $A$ is the wall amplitude, $\lambda=L_{\mathrm{sin}}/n$ is the wavelength, $n$ is the number of sinusoidal waves in the active section, and $\phi$ is the phase offset between the upper and lower walls. The wall amplitude is smoothly modulated by the envelope function $E(\xi)$, defined as $E(\xi)=\frac{1}{2}\left[1-\cos\left(\pi\xi/L_r\right)\right]$ for $0\le\xi<L_r$, $E(\xi)=1$ for $L_r\le\xi\le L_{\mathrm{sin}}-L_r$, and $E(\xi)=\frac{1}{2}\left[1-\cos\left(\pi(L_{\mathrm{sin}}-\xi)/L_r\right)\right]$ for $L_{\mathrm{sin}}-L_r<\xi\le L_{\mathrm{sin}}$, where the ramp length is $L_r=\lambda/2$. This envelope satisfies $E(0)=E(L_{\mathrm{sin}})=0$ and $E'(0)=E'(L_{\mathrm{sin}})=0$, ensuring that the sinusoidal walls connect to the straight inlet and outlet sections with continuous position and slope. Consequently, geometric discontinuities at $x=500\,\mu\mathrm{m}$ and $x=2900\,\mu\mathrm{m}$ are eliminated.

\subsection{Governing Equations}

The flow field is governed by the conservation equations for mass, momentum, and scalar transport. For an incompressible fluid, the continuity and momentum equations are given by
\begin{equation}
\nabla \cdot \mathbf{u} = 0, \quad\quad \rho
\left(
\frac{\partial \mathbf{u}}{\partial t}
+
\mathbf{u}\cdot\nabla \mathbf{u}
\right)
=
-\nabla p
+
\nabla \cdot \boldsymbol{\tau},
\label{eq:continuity_momentum}
\end{equation}
where $\mathbf{u}$ denotes the velocity vector, $\rho$ is the fluid density, $p$ is the pressure, and $\boldsymbol{\tau}$ is the extra-stress tensor. For a generalized Newtonian fluid, the extra-stress tensor is defined as $\boldsymbol{\tau}
=
2\mu(\dot{\gamma})\mathbf{D},$
where $\mathbf{D}$ is the rate-of-deformation tensor,
$\mathbf{D}
=
\frac{1}{2}
\left(
\nabla \mathbf{u}
+
\nabla \mathbf{u}^{T}
\right),$
and the magnitude of the shear rate is given by
$
\dot{\gamma}
=
\sqrt{2\mathbf{D}:\mathbf{D}}.$
The working fluid was modeled as an incompressible shear-thinning fluid described by the Carreau--Yasuda constitutive model. We evaluated the apparent viscosity using the Carreau--Yasuda constitutive relation~\cite{Carreau1972,Yasuda1981,bird1987dynamics},
\begin{equation}
\mu_{\text{app}}
=
\mu_{\infty}
+
(\mu_{0}-\mu_{\infty})
\left[
1+(\lambda_c \dot{\gamma})^{a}
\right]^{\frac{n_c-1}{a}},
\label{eq:carreau_yasuda}
\end{equation}
where $\mu_0$ and $\mu_{\infty}$ denote the zero-shear and infinite-shear viscosities, respectively, $\lambda_c$ represents the characteristic time constant governing the onset of shear-thinning, $n_c$ is the power-law index controlling the degree of shear-thinning, and $a$ is the Yasuda transition parameter controlling the breadth of the transition region between the Newtonian plateaus. Table~\ref{tab:carreau_params} summarizes the fluid properties used in the simulations. The Carreau-Yasuda model is utilized here because it accurately captures the non-Newtonian, shear-thinning behavior of blood by smoothly bounding its viscosity between physiological extremes~\cite{weddell2015hemodynamic}. The specific parameters reflect established empirical measurements for human blood: the zero-shear viscosity ($\mu_0$ = 0.056 Pa s) represents the high flow resistance caused by red blood cell aggregation, while the infinite-shear viscosity ($\mu_\infty$ = 0.0035 Pa s) accounts for cellular deformation and flow alignment at high shear rates~\cite{lai2024review}. The characteristic time constant ($\lambda_c$ = 1.902 s), power-law index ($n_c$ = 0.22), and Yasuda parameter ($a$ = 1.25) are constants tuned to govern the onset and curvature of this transition, ensuring the mathematical simulation strictly aligns with physical rheological data~\cite{kumar2022influence}.
\begin{table}[h!]
\centering
\caption{Carreau--Yasuda fluid parameters used in the simulations}
\label{tab:carreau_params}
\begin{tabular}{llcc}
\toprule
\textbf{Parameter} &
\textbf{Description} &
\textbf{Value} &
\textbf{Unit} \\
\midrule
$\rho$ &
Density &
1000 &
kg\,m$^{-3}$ \\

$\mu_0$ &
Zero-shear viscosity &
0.056 &
Pa\,s \\

$\mu_\infty$ &
Infinite-shear viscosity &
0.0035 &
Pa\,s \\

$\lambda_c$ &
Time constant &
1.902 &
s \\

$n_c$ &
Power-law index &
0.22 &
--- \\

$a$ &
Yasuda parameter &
1.25 &
--- \\

$D$ &
Molecular diffusivity &
$1 \times 10^{-9}$ &
m$^2$\,s$^{-1}$ \\
\bottomrule
\end{tabular}
\end{table}

To quantify species transport, we solved the transient
convection--diffusion equation,
\begin{equation}
\frac{\partial C}{\partial t}
+
\mathbf{u}\cdot\nabla C
=
D\nabla^{2}C,
\label{eq:concentration}
\end{equation}
where \(C\) denotes the normalized scalar concentration,
\(\mathbf{u}\) is the velocity field, and \(D\) is the molecular
diffusivity. We quantified the degree of mixing using the mixing index
(\(MI\)), defined at the outlet cross-section as
\begin{equation}
MI
=
1-\frac{\sigma}{\sigma_0},
\qquad
\sigma
=
\sqrt{\frac{1}{N}
\sum_{i=1}^{N}
\left(C_i-\bar{C}\right)^2},
\label{eq:MI}
\end{equation}
where \(\sigma\) is the standard deviation of the outlet concentration
distribution, \(\sigma_0\) is the corresponding standard deviation for
the unmixed inlet condition, \(N\) is the number of sampling points,
and \(\bar{C}\) is the mean concentration. We evaluated the outlet mixing index (\(MI\)) by sampling the concentration field at 500 uniformly distributed points across the outlet cross-section (\(x = 3400\,\mu\mathrm{m}\)). We define the pressure drop, \(\Delta p\), as the difference between the cross-sectionally averaged pressures at \(x = 500\,\mu\mathrm{m}\) and \(x = 2900\,\mu\mathrm{m}\), i.e., $\Delta p = \bar{p}\left(x = 500\,\mu\mathrm{m}\right) -
\bar{p}\left(x = 2900\,\mu\mathrm{m}\right)$, where \(\bar{p}\) denotes the pressure averaged over the channel cross-section.

We prescribed a uniform inlet velocity of
\(u_{\mathrm{in}}=0.005~\mathrm{m\,s^{-1}}\). Based on the channel
half-height (\(h=100~\mu\mathrm{m}\)), the Reynolds number and Péclet number remained fixed throughout the study, with
\(
Re=\rho u_{\mathrm{in}}h/\mu_0 \approx 0.009
\)
and
\(
Pe=u_{\mathrm{in}}h/D=500
\).
The low Reynolds number ensures laminar creeping flow, whereas the moderate Péclet number indicates convection-dominated streamwise transport with transverse mixing governed by molecular diffusion. Consequently, the sinusoidal converging--diverging geometry enhances mixing by repeatedly deforming the concentration field, increasing the interfacial area and concentration gradients that drive molecular diffusion.

\subsection{Numerical Solution Procedure}

We solved the governing equations using the finite-volume solver \texttt{rheoFoam} within the RheoTool framework~\cite{Pimenta2017}, which is built on Open Field Operation and Manipulation (OpenFOAM) version~9. The computational domain was discretized using structured two-dimensional finite-volume meshes. The transient simulations were advanced until the flow and concentration fields attained a statistically stationary state, from which the pressure-drop and concentration distributions were obtained. Pressure--velocity coupling was achieved using the merged \textit{Pressure-Implicit with Splitting of Operators} and \textit{Semi-Implicit Method for Pressure-Linked Equations} (PIMPLE) algorithm. Time integration was performed using a first-order implicit Euler scheme, while spatial discretization employed second-order Gaussian schemes. To reduce numerical diffusion, the \textit{Convergent and Universally Bounded Interpolation Scheme for the Treatment of Advection} (CUBISTA) was used for the discretization of scalar and constitutive convective terms, whereas corrected Gaussian schemes were applied to the diffusive terms. The pressure equation was solved using the \textit{Preconditioned Conjugate Gradient} (PCG) solver with a \textit{Diagonal Incomplete Cholesky} (DIC) preconditioner. The momentum, scalar transport, and constitutive equations were solved using the \textit{Preconditioned Bi-Conjugate Gradient} (PBiCG) solver with a \textit{Diagonal Incomplete Lower--Upper} (DILU) preconditioner. A fixed time step of $\Delta t = 10^{-3}$ was employed while maintaining the Courant number below unity throughout the simulations. Computations were continued until both the pressure drop across the sinusoidal section and the outlet mixing index reached convergence. The upper and lower inlet streams were assigned normalized scalar concentrations of $C=1$ and $C=0$, respectively. No-slip and no-flux boundary conditions were imposed at the channel walls, while zero-gradient boundary conditions were prescribed at the outlet for all transported variables except pressure, for which a fixed reference value was specified. The outlet mixing index was evaluated from the converged concentration field, whereas the pressure drop was calculated from the pressure difference across the sinusoidal section.

\subsection{Numerical Verification}
We solved the governing equations using \texttt{rheoFoam}, a specialized solver within the open-source RheoTool~\cite{Pimenta2017} framework. Since its foundational validation for viscoelastic flows~\cite{Pimenta2017}, \texttt{rheoFoam} has been extensively verified against alternative numerical methods for generalized Newtonian, thixotropic, and elastoviscoplastic fluids~\cite{CastilloSanchez2022, Fernandes2019}. The solver's robustness in capturing severe stress gradients is well documented across complex microfluidic geometries, including abrupt converging--diverging contractions~\cite{Rodrigues2023, CastilloSanchez2022}. Here, we mainly focused on verifying the numerical discretization adopted for the present sinusoidal converging--diverging microchannels. We performed mesh and time-step independence studies using a representative geometry with
\(A/h=0.5\), \(\phi=120^{\circ}\), and \(n=4\). We progressively refined the structured mesh and reduced the time step while monitoring the outlet mixing index (\(MI\)) and the pressure drop across the sinusoidal section (\(\Delta p\)). As summarized in Table~\ref{tab:verification}, refining the mesh from Mesh~3 to Mesh~4 changed \(MI\) by less than \(0.1\%\) and \(\Delta p\) by less than \(0.05\%\). Likewise, reducing the time step from \(10^{-3}\) to \(10^{-4}\) changed \(MI\) by less than \(0.1\%\) and \(\Delta p\) by less than \(0.04\%\). These negligible differences confirm mesh and temporal independence. Accordingly, we adopted Mesh~3 and a time step of \(10^{-3}\) for all subsequent simulations. Each simulation was continued until both \(MI\) and \(\Delta p\) reached steady values, ensuring numerically converged solutions.
\begin{table}[htbp]
\centering
\caption{Mesh and time-step independence studies for the representative geometry (\(A/h=0.5\), \(\phi=120^{\circ}\), and \(n=4\)).}
\label{tab:verification}
\renewcommand{\arraystretch}{1.15}
\begin{tabular}{llccccc}
\hline
\textbf{Study} & \textbf{Case} & \textbf{Nodes} & \textbf{Min. Size (m)} & \textbf{Max. Size (m)} & $MI$ & $\Delta p$ \textbf{(Pa)} \\
\hline
\multirow{4}{*}{Mesh}
& Mesh 1 & 29\,069  & \(3\times10^{-6}\) & \(5\times10^{-6}\) & 0.227 & 27.751 \\
& Mesh 2 & 45\,135  & \(2\times10^{-6}\) & \(4\times10^{-6}\) & 0.226 & 27.780 \\
& Mesh 3 & 80\,240  & \(1\times10^{-6}\) & \(3\times10^{-6}\) & 0.226 & 27.792 \\
& Mesh 4 & 178\,770 & \(8\times10^{-7}\) & \(1\times10^{-6}\) & 0.226 & 27.777 \\
\hline
\multirow{4}{*}{Time Step}
& \(\Delta t=10^{-1}\) & 80\,240 & \(1\times10^{-6}\) & \(3\times10^{-6}\) & 0.223 & 25.824 \\
& \(\Delta t=10^{-2}\) & 80\,240 & \(1\times10^{-6}\) & \(3\times10^{-6}\) & 0.226 & 27.805 \\
& \(\Delta t=10^{-3}\) & 80\,240 & \(1\times10^{-6}\) & \(3\times10^{-6}\) & 0.226 & 27.804 \\
& \(\Delta t=10^{-4}\) & 80\,240 & \(1\times10^{-6}\) & \(3\times10^{-6}\) & 0.226 & 27.792 \\
\hline

\end{tabular}
\end{table}

\section{Surrogate-Modeling Methodology}
\label{sec:methods}
High-fidelity CFD simulations provide accurate predictions of micromixer performance but make exhaustive exploration of the design space computationally expansive. To enable rapid design exploration, we developed surrogate models that relate the geometric design parameters to the mixing index ($MI$) and pressure drop ($\Delta p$). We trained and evaluated multiple regression models using the CFD-generated dataset and assessed their predictive capability through cross-validation, internal testing, and an independent holdout dataset. We subsequently employed the best-performing surrogate for response-surface analysis, Pareto-front construction, and multi-objective optimization to identify micromixer geometries that achieve an optimal balance between mixing performance and pressure drop.

\subsection{Dataset and Feature Representation}
\label{sec:features}
The CFD database comprises 218 simulations, including 169 cases for surrogate-model development and 49 independently generated cases reserved for external validation. The primary dataset consists of 168 sinusoidal microchannel geometries and one straight-channel reference case is assigned $\sin\phi = \cos\phi = 0$ and $\lambda = 0$, since the phase offset and wave count are undefined in the absence of wall modulation. Sinusoidal geometries span \(A/h=\{0.3,0.5,0.7,0.9\}\), \(\phi=\{0^\circ,30^\circ,60^\circ,90^\circ,120^\circ,150^\circ,180^\circ\}\), and \(n=\{2,3,4,5,6,8\}\), corresponds to wavelengths \(\lambda=L_{\mathrm{sin}}/n=\{1200,800,600,480,400,300\}\,\mu\mathrm{m}\).

We represented each geometry by the feature vector \(\mathbf{x}=[A/h,\sin\phi,\cos\phi,\lambda]^{\mathsf T}\). The trigonometric encoding preserves the periodicity of the phase offset and eliminates the discontinuity at \(0^\circ\equiv360^\circ\), while replacing the discrete wave count \(n\) with the continuous wavelength \(\lambda\) facilitates interpolation between neighboring geometries. Across the primary dataset, the mixing index ranges from \(0.21\) to \(0.38\), whereas the pressure drop spans \(15.68\) to \(817.27~\mathrm{Pa}\). We therefore trained the pressure-drop surrogate using \(\log(\Delta p)\). The external validation set contains 49 unseen geometries, including intermediate \(n=7\) configurations, providing a stringent test of interpolation capability.

\subsection{Surrogate Models}
\label{sec:surrogates}
We used the CFD dataset to construct surrogate models that approximate the mapping between the geometric parameters and the corresponding performance metrics, namely the mixing index (\(MI\)) and pressure drop (\(\Delta p\)). To assess the influence of model complexity, we evaluated four regression algorithms with increasing representational flexibility: ordinary least-squares linear regression, polynomial ridge regression (PRR), Gaussian process regression (GPR), and Extreme Gradient Boosting (XGBoost). We adopted ordinary least-squares linear regression as the baseline model. To account for nonlinear feature interactions, we developed polynomial ridge regression models using second- and third-degree polynomial feature expansions. We selected the regularization parameter from \(\alpha\in\{10^{-6},\,10^{-4},\,10^{-2},\,1\}\). Cross-validation identified the third-degree expansion as optimal for both targets, with \(\alpha=10^{-4}\) for \(MI\) and \(\alpha=10^{-2}\) for \(\log(\Delta p)\).

To provide a probabilistic nonlinear surrogate, we trained separate GPR models for \(MI\) and \(\log(\Delta p)\) using exact Gaussian process inference~\cite{Rasmussen2006}. We adopted a Mat\'{e}rn-\(\tfrac{5}{2}\) covariance function with automatic relevance determination (ARD) and an additive white-noise term,
\begin{equation}
k\!\left(\widetilde{\mathbf{x}}_{i},\widetilde{\mathbf{x}}_{j}\right)
=
\sigma_f^2
\left(1+\sqrt{5}\,r_{ij}+\frac{5r_{ij}^{2}}{3}\right)
\exp\!\left(-\sqrt{5}\,r_{ij}\right)
+\sigma_n^2\delta_{ij},
\quad
r_{ij}
=
\sqrt{\sum_{d=1}^{4}
\dfrac{\left(\widetilde{x}_{i,d}-\widetilde{x}_{j,d}\right)^2}{\ell_d^2}},
\label{eq:matern_kernel}
\end{equation}

where \(\sigma_f^2\) is the signal variance, \(\ell_d\) is the ARD length scale for feature \(d\), \(\sigma_n^2\) is the white-noise variance, and \(\delta_{ij}\) is the Kronecker delta. Because the CFD responses are deterministic, we used the white-noise term only as a numerical nugget to improve covariance-matrix conditioning and absorb minor residual variation arising from iterative convergence; we did not interpret it as discretization error. We standardized each target before training and estimated the kernel hyperparameters by maximizing the log marginal likelihood using the L-BFGS-B optimizer. During validation, we used three optimizer restarts and screened initial white-noise levels of \(10^{-8}\), \(10^{-6}\), and \(10^{-4}\), selecting \(10^{-6}\) for \(MI\) and \(10^{-8}\) for \(\log(\Delta p)\). We increased the number of optimizer restarts to ten when fitting the final Gaussian process (GP) models. Under the Gaussian posterior assumption, the GP posterior mean and standard deviation were used to construct approximate \(95\%\) predictive intervals. For the mixing index, the interval was computed as
\(\hat{y} \pm 1.96\,\hat{\sigma}\). For the pressure drop, which was modeled in logarithmic space, the interval was obtained by back-transforming the bounds,
\(\exp(\hat{\mu}_{\log} \pm 1.96\,\hat{\sigma}_{\log})\), yielding asymmetric predictive intervals in physical units. The GP posterior variance was used solely to quantify predictive uncertainty.

We further considered XGBoost as a nonlinear tree-ensemble benchmark~\cite{Chen2016}. Cross-validation selected 600 boosting iterations for both targets. For the \(MI\) model, the optimal hyperparameters were a learning rate of 0.05, maximum tree depth of 2, minimum child weight of 1, subsampling fraction of 0.9, and an \(\ell_2\)-regularization strength of 1.0. For the \(\log(\Delta p)\) model, the optimal configuration consisted of a learning rate of 0.02, maximum tree depth of 3, minimum child weight of 1, subsampling fraction of 0.9, and the same \(\ell_2\)-regularization strength. We interpreted the trained XGBoost models using Tree SHapley Additive exPlanations (TreeSHAP) values~\cite{Lundberg2017}. For the pressure surrogate, we computed SHapley Additive exPlanations (SHAP) values in the transformed ($\log(\Delta p)$) space. Because the phase angle was encoded using the pair $(\sin\phi,\cos\phi)$, we combined their mean absolute SHAP values to quantify the overall importance of the phase variable. This grouped importance represents the combined contribution of the two encoded features and should not be interpreted as a SHAP interaction value. We interpret SHAP values solely as measures of model attribution rather than evidence of physical causality.

We assessed surrogate performance using the coefficient of determination (\(R^2\)), root-mean-square error (\(RMSE\)), mean
absolute error (\(MAE\)), maximum absolute error
(\(\epsilon_{\max}\)), and prediction bias. For CFD targets \(y_i\),
surrogate predictions \(\hat{y}_i\), target mean \(\bar{y}\), and
\(N\) samples, these metrics are defined as
\(R^2 = 1-\sum_i (y_i-\hat{y}_i)^2/\sum_i (y_i-\bar{y})^2\),
\(RMSE = \sqrt{\sum_i (y_i-\hat{y}_i)^2/N}\),
\(MAE = \sum_i |y_i-\hat{y}_i|/N\),
\(\epsilon_{\max} = \max_i |y_i-\hat{y}_i|\), and
\(\mathrm{Bias} = \sum_i (\hat{y}_i-y_i)/N\).
The \(R^2\) score quantifies the fraction of variance explained by the
surrogate, \(RMSE\) and \(MAE\) measure the average prediction error,
\(\epsilon_{\max}\) identifies the worst-case deviation, and
\(\mathrm{Bias}\) detects systematic over- or under-prediction.

\subsection{Validation and Model-Selection Protocol}
\label{sec:validation}

We established a unified validation framework to ensure an unbiased comparison of surrogate-model performance. The framework defines the data partitioning, hyperparameter optimization, cross-validation procedure, and evaluation metrics used for model selection. 
\begin{table}[htbp]
    \centering
    \caption{Validation hierarchy used for surrogate development.}
    \label{tab:validation_framework}
    \begin{tabular}{lll}
        \hline
        Assessment tier & Cases & Role \\
        \hline
        Training & 119 & Model fitting \\
        Internal validation & 25 & Hyperparameter selection \\
        Internal test & 25 & Same-fit performance assessment \\
        Five-fold cross-validation & 168 (5 folds)
                                   & Post-selection stability check \\
        External holdout & 49 & Independent unseen-design assessment \\
        \hline
    \end{tabular}
\end{table}
We generated the partition in Table~\ref{tab:validation_framework} deterministically using greedy maximin selection on the standardized feature space. This procedure selected mutually distant sinusoidal geometries to construct a space-filling holdout sample, which we subsequently divided into disjoint validation and test sets. We selected the model hyperparameter by minimizing the validation RMSE, using the MAE as a secondary criterion when multiple configurations produced similar RMSE values. Throughout model selection, we excluded the internal-test and external-holdout sets from training and hyperparameter tuning. After selecting the optimal hyperparameter, we performed five-fold cross-validation on the 168 sinusoidal geometries to assess model robustness across different data partitions. Because we kept the hyperparameter fixed during cross-validation, this step evaluated model stability rather than serving as an additional model-selection stage.

To compare the four surrogate models, we first trained each model using the same 119-case training set and evaluated its performance on the internal-test and external-holdout sets. After selecting the best-performing configuration, we retrained the final model using all 169 primary cases and performed a final evaluation on the 49 unseen external cases. We used this final surrogate for all subsequent analyses, including parity and error assessment, response-surface construction, feature-importance analysis, and multi-objective optimization.

\section{Multi-Objective Optimization Framework}
\label{sec:optimisation}
\subsection{Design Variables, Feasible Domain, and Objectives}
\label{sec:mop}

We formulated the surrogate-based multi-objective optimization problem to maximize the GP-predicted outlet mixing index while simultaneously minimizing the GP-predicted pressure drop. We defined the design vector as \(\mathbf{z}=[A/h,\;\phi,\;n]\), subject to \(0.3\le A/h\le0.9\), \(0^\circ\le\phi\le180^\circ\), and \(n\in\{2,3,4,5,6,7,8\}\). We excluded \(A/h=0\) because the wall modulation vanishes in this limit, reducing the geometry to a straight channel in which neither the phase shift $\phi$ nor the number of sinusoidal periods $n$ influences the flow.

We treated \(n\) as an integer design variable because it specifies the number of complete sinusoidal periods over the fixed active-section length. Accordingly, we performed independent NSGA-II optimizations for each admissible value of \(n\), thereby avoiding a mixed-integer optimization problem. Optimization for \(n=7\) also allowed us to assess the interpolation capability of the GP surrogate because this configuration is located between the training levels \(n=6\) and \(n=8\).

We coupled the final-refit GP surrogates with NSGA-II by solving
\begin{equation}
\min_{\mathbf{z}\in\mathcal{Z}}
\left[
-\widehat{MI}_{\mathrm{GP}}(\mathbf{z}),\;
\exp\!\left(\widehat{y}_{\Delta p,\mathrm{GP}}(\mathbf{z})\right)
\right],
\label{eq:multiobjective_problem}
\end{equation}
where \(\widehat{MI}_{\mathrm{GP}}\) and \(\widehat{y}_{\Delta p,\mathrm{GP}}\) denote the GP predictions of the mixing index and \(\log(\Delta p)\), respectively. We applied an exponential transformation to the prediction of the pressure to recover the pressure drop (in Pa), allowing us to optimize the physical hydraulic performance together with the mixing efficiency.

\subsection{Pareto-Front Construction and Knee Selection}
\label{sec:pareto_knee}

We extracted the CFD Pareto front from the 168 sinusoidal primary simulations using the non-dominance criterion~\cite{Miettinen1999} and constructed the GP--NSGA-II front from the globally non-dominated surrogate solutions. We then identified the geometric knee, which represents the best trade-off between the competing objectives, as the point with the maximum perpendicular distance from the chord connecting the low-pressure and high-mixing extremes of the Pareto front,
\begin{equation}
\mathbf{z}_{\mathrm{knee}}
=
\arg\max_{\mathbf{z}\in\mathcal{P}}
d_{\perp}
\!\left(
\widetilde{\mathbf{p}}(\mathbf{z}),
\overline{\widetilde{\mathbf{p}}_{\min}\widetilde{\mathbf{p}}_{\max}}
\right),
\label{eq:knee}
\end{equation}
where we normalized the CFD and GP--NSGA-II fronts using common objective bounds obtained from their union.

As a sensitivity check, we also identified the equal-weight ideal-point solution,
\begin{equation}
\mathbf{z}_{\mathrm{ideal}}
=
\arg\min_{\mathbf{z}\in\mathcal{P}}
\sqrt{
\widetilde{\Delta p}^{\,2}
+
\left(1-\widetilde{MI}\right)^2},
\label{eq:ideal_compromise}
\end{equation}
and, for a prescribed pressure limit \(\varepsilon\), determined the pressure-constrained optimum from
\begin{equation}
\mathbf{z}_{\varepsilon}
=
\arg\max_{\mathbf{z}\in\mathcal{P}}
\widehat{MI}_{\mathrm{GP}}(\mathbf{z})
\quad
\text{subject to}
\quad
\widehat{\Delta p}_{\mathrm{GP}}(\mathbf{z})\le\varepsilon.
\label{eq:pressure_constrained}
\end{equation}

\subsection{GP-Assisted Multi-Objective Optimization and CFD Validation}
\label{sec:nsga2}

We coupled the final-refit GP surrogates with NSGA-II~\cite{Deb2002} using the \texttt{pymoo} framework~\cite{BlankDeb2020}. For each integer wave count \(n\), we performed three independent optimization runs with random seeds 11, 42, and 101. 
Each run used a population of 120, initialized by uniform random sampling, and
evolved over 200 generations, with simulated binary crossover (crossover
probability 0.90, crossover distribution index 15), polynomial mutation
(mutation distribution index 20), and per-generation duplicate elimination. We then combined the non-dominated solutions obtained from all optimization runs and wave counts, eliminated near-duplicate designs, and applied a global non-dominance filter to obtain the final GP--NSGA-II Pareto front.\\

We reconstructed the GP--NSGA-II knee geometry and evaluated it using an independent OpenFOAM simulation with the same governing equations, numerical settings, boundary conditions, convergence criteria, and post-processing procedures used for the primary CFD database. To ensure an unbiased assessment, we excluded this confirmation case from surrogate training, hyperparameter selection, and optimization. We then compared the GP predictions with the corresponding CFD values of \(MI\) and \(\Delta p\), and report the corresponding discrepancies in Sec.~\ref{sec:results_discussion}.

\section{Results and Discussion}
\label{sec:results_discussion}

The present flow remains in the creeping-flow regime (\(Re\approx0.009\)), where viscous forces dominate inertia and turbulence is absent. Under these conditions, sinusoidal wall modulation cannot enhance mixing through inertial instabilities or turbulent dispersion. Instead, mixing originates from the coupled interaction between geometric deformation, shear-thinning rheology, and molecular diffusion. Figure~\ref{fig:contours} presents the concentration, velocity-gradient, and apparent-viscosity contours for a straight channel and two sinusoidal microchannels with different degrees of geometric modulation, illustrating the physical mechanisms governing the scalar transport and flow field.

\subsection{Flow Physics of Mixing Enhancement}

\subsubsection{Effect of Geometric Modulation on Scalar Transport}

\begin{figure}[t]
    \centering
    \includegraphics[width=\linewidth]{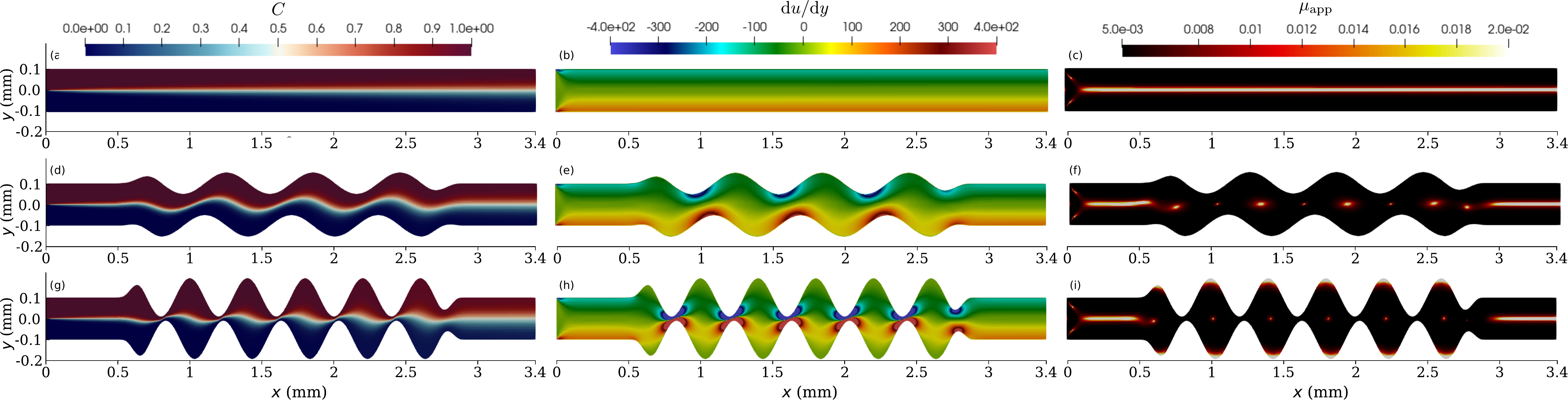}
    \caption{Contours of concentration distribution, velocity gradient, and apparent viscosity for (a--c) the straight channel, (d--f) the sinusoidal channel with \(A/h=0.5\), \(\phi=90^\circ\), and \(n=4\), and (g--i) the sinusoidal channel with \(A/h=0.9\), \(\phi=150^\circ\), and \(n=6\).}
    \label{fig:contours}
\end{figure}

The concentration contours in Fig.~\ref{fig:contours}(a,d,g) demonstrate that geometric modulation alters the scalar transport mechanism. In the straight channel (Fig.~\ref{fig:contours}(a)), the concentration interface remains nearly sharp throughout the domain because the streamlines remain almost parallel under creeping-flow conditions. Consequently, molecular diffusion acts across a relatively small interfacial area, resulting in slow transverse mass transport and limited outlet homogenization. Introducing sinusoidal wall modulation continuously perturbs this otherwise stable interface. As the fluid traverses successive converging--diverging sections, alternating acceleration and deceleration repeatedly stretch and elongate the
concentration interface, increasing both its length and curvature. This geometric deformation continuously generates new concentration gradients normal to the interface, thereby enlarging the effective diffusion area without altering the molecular diffusivity. Repeated interface stretching continually increases local concentration gradients, thereby enhancing molecular diffusion and accelerating scalar homogenization along the channel.
Increasing the wall modulation further intensifies this mechanism. Compared with the moderately modulated geometry (\(A/h=0.5\), \(\phi=90^\circ\), \(n=4\); Fig.~\ref{fig:contours}(d)), the strongly modulated channel (\(A/h=0.9\), \(\phi=150^\circ\), \(n=6\); Fig.~\ref{fig:contours}(g)) produces significantly greater interface elongation and repeated curvature during each contraction--expansion cycle. The cumulative stretching generated over successive sinusoidal units therefore increases scalar-gradient production and promotes substantially higher outlet mixing. These observations indicate that, in the absence of inertial mixing, sinusoidal geometric modulation enhances transport primarily by increasing interfacial area and strengthening diffusion-driven homogenization.

\subsubsection{Shear-Rate Amplification and Shear-Thinning Response}

The enhanced scalar transport directly couples with the non-Newtonian rheology of the Carreau--Yasuda fluid. The velocity-gradient contours in Fig.~\ref{fig:contours}(b,e,h) show that each converging section generates localized regions high velocity gradient regions. These regions consistently occur near the contraction throats, whereas the expanding sections allow partial relaxation of the deformation field. This kinematic response produces a corresponding rheological response through the Carreau--Yasuda constitutive model. As shown in Fig.~\ref{fig:contours}(c,f,i), the regions of maximum velocity gradient coincide with the lowest apparent viscosity because the local shear rate exceeds the transition between the Newtonian plateau and the shear-thinning regime. Consequently, the apparent viscosity decreases within the contraction regions while remaining comparatively larger in the low-shear expansion zones. The spatial redistribution of viscosity further amplifies interfacial deformation. The reduced apparent viscosity lowers the local viscous resistance, allowing the fluid to accelerate more readily through successive contractions and increasing the stretching rate of the concentration interface. Consequently, geometric modulation and shear-thinning rheology act synergistically: stronger contractions produce larger velocity gradients, the elevated shear rates reduce the apparent viscosity, and the resulting reduction in flow resistance promotes even greater interface deformation. This positive coupling explains the substantial improvement in mixing observed as the wall amplitude increases.

\subsubsection{Origin of the Hydraulic Penalty}

Although stronger geometric modulation enhances scalar transport, it simultaneously increases the hydraulic cost required to sustain the prescribed flow rate. Figure~\ref{fig:contours}(b,e,h) shows that increasing the wall amplitude progressively concentrates large velocity gradients within the contraction regions. Maintaining a constant volumetric flow rate through these narrower passages therefore requires larger pressure gradients to overcome the increased viscous resistance associated with repeated acceleration and deceleration of the fluid. The shear-thinning response partially alleviates this hydraulic penalty by reducing the apparent viscosity within the high-shear regions, as illustrated in Fig.~\ref{fig:contours}(c,f,i). However, this rheological benefit cannot fully compensate for the increased geometric confinement imposed by larger wall amplitudes and repeated contraction--expansion cycles. Consequently, viscous dissipation accumulates along the channel, producing progressively larger pressure losses as the geometric modulation becomes stronger.

The simultaneous enhancement of mixing and pressure drop therefore reflects two competing consequences of the same physical mechanism. The velocity gradients responsible for stretching the concentration interface and accelerating molecular diffusion also increase viscous resistance within the flow. As a result, geometries that maximize scalar transport inevitably incur larger pressure drop. This intrinsic competition between transport enhancement and flow resistance defines the fundamental design trade-off investigated throughout the remainder of this work and motivates the subsequent surrogate-assisted multi-objective optimization. The representative flow fields in Fig.~\ref{fig:contours} reveal the mechanisms governing mixing enhancement in sinusoidal converging--diverging microchannels. However, they cannot fully capture the coupled effects of wall-amplitude ratio, phase offset, and wave count across the design space. To overcome the computational cost of exhaustive high-fidelity CFD simulations, we develop surrogate models from the CFD database to rapidly predict the mixing index and pressure drop. The following subsection evaluates their predictive performance.

\subsection{Surrogate Model Development and Validation}

The nonlinear coupling between geometric modulation, shear-thinning rheology, and scalar transport produces a complex relationship between the design variables and the corresponding mixing and pressure drop behavior. Although the CFD database accurately resolves these interactions, its computational cost limits exhaustive exploration of the multidimensional design space. We therefore developed surrogate models to rapidly approximate the mappings between the geometric parameters and the outlet mixing index (\(MI\)) and pressure drop (\(\Delta p\)). This section compares the predictive performance of the candidate surrogate models, evaluates their robustness on unseen geometries, and examines their generalization capability.

\subsubsection{Selection of the Optimal Surrogate Model}

\begin{figure}[t]
    \centering
\includegraphics[width=0.8\linewidth]{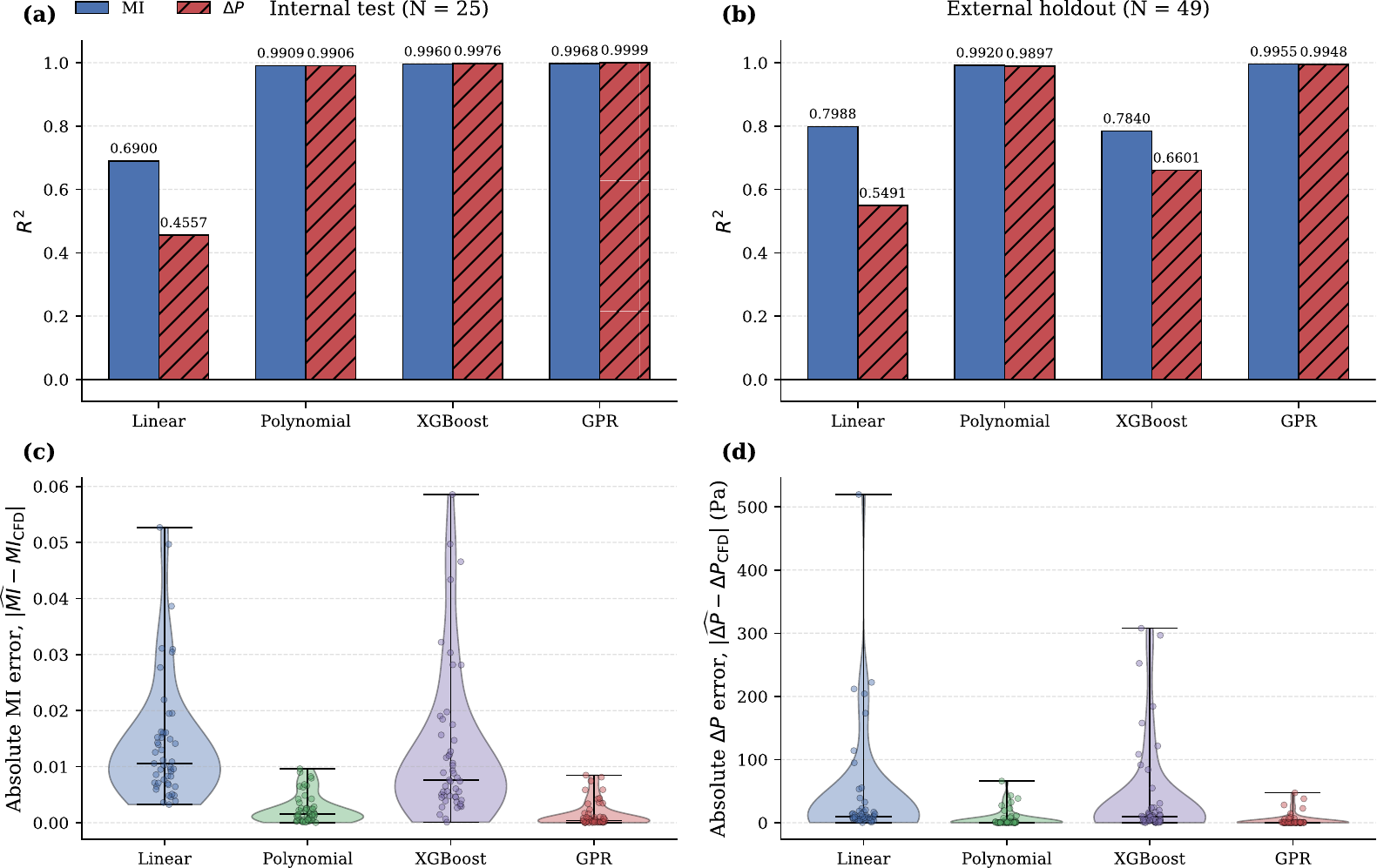}
    \caption{Comparison of surrogate-model performance for predicting the mixing index (\(MI\)) and pressure drop (\(\Delta p\)) in sinusoidal converging--diverging microchannels. (a,b) Coefficient of determination (\(R^2\)) for Linear Regression, Polynomial Regression, XGBoost, and Gaussian Process Regression (GPR) on the internal test set (\(N=25\)) and external holdout set (\(N=49\)), respectively. (c,d) Violin plots of the absolute prediction errors for \(MI\) and \(\Delta p\) on the external holdout set.}
    \label{fig:surrogate_comparison}
\end{figure}

We evaluated the predictive performance of Linear Regression (LR), Polynomial Ridge Regression (PRR), XGBoost, and Gaussian Process Regression (GPR) to identify the most suitable surrogate model for subsequent optimization. Figure~\ref{fig:surrogate_comparison} compares the predictions of the four models, and Table~\ref{tab:internal_test_performance} summarizes their quantitative performance metrics on the internal test dataset.
\begin{table}[h]
    \small
    \centering
    \caption{Internal-test performance of surrogate models}
    \label{tab:internal_test_performance}
    \begin{minipage}[t]{0.48\textwidth}
        \centering
        Mixing Index ($MI$)\\[4pt]
        \begin{tabular}{lcccccc}
            \toprule
            Model & $R^2$ & RMSE & MAE & $\epsilon_{\max}$ & Bias \\
            \midrule
            Linear Reg.  & 0.6900 & 0.02172 & 0.01696 & 0.05177 & $-0.00495$ \\
            Poly.\ Ridge & 0.9909 & 0.00371 & 0.00249 & 0.01145 & $-0.00073$ \\
            XGBoost      & 0.9960 & 0.00248 & 0.00174 & 0.00689 & $-0.00050$ \\
            GPR          & 0.9968 & 0.00222 & 0.00085 & 0.01052 & $-0.00027$ \\
            \bottomrule
        \end{tabular}
    \end{minipage}%
    \hfill
    \begin{minipage}[t]{0.48\textwidth}
        \centering
        Pressure Drop ($\Delta p$ in Pa)\\[4pt]
        \begin{tabular}{lcccccc}
            \toprule
            Model & $R^2$ & RMSE & MAE & $\epsilon_{\max}$ & Bias \\
            \midrule
            Linear Reg.  & 0.4557 & 104.69 & 32.84 & 513.52 & $-19.45$ \\
            Poly.\ Ridge & 0.9906 & 13.73  & 5.39  & 57.75  & $-0.85$  \\
            XGBoost      & 0.9976 & 6.97   & 2.69  & 31.54  & $-1.06$  \\
            GPR          & 0.9999 & 0.54   & 0.29  & 1.82   & $0.15$   \\
            \bottomrule
        \end{tabular}
    \end{minipage}
\end{table}

Linear regression consistently exhibits the weakest predictive capability for both objectives. As shown in Fig.~\ref{fig:surrogate_comparison}(a), it achieves only \(R^2=0.6900\) for the mixing index and \(R^2=0.4557\) for the pressure drop. The corresponding RMSE values listed in Table~\ref{tab:internal_test_performance} are \(2.17\times10^{-2}\) for \(MI\) and \(104.69~\mathrm{Pa}\) for \(\Delta p\), nearly two orders of magnitude larger than those obtained using the nonlinear surrogate models. These results demonstrate that additive linear relationships cannot adequately represent the coupled influence of wall amplitude ratio, phase offset, and wavelength on the transport characteristics. Geometric modulation simultaneously alters the local shear rate, apparent viscosity, and concentration-field, producing highly nonlinear interactions that extend beyond the capability of a linear approximation.

Introducing higher-order feature interactions substantially improves predictive accuracy. The third-degree PRR model increases the coefficient of determination to \(R^2=0.9909\) for \(MI\) and \(R^2=0.9906\) for \(\Delta p\), while reducing the RMSE to \(3.71\times10^{-3}\) and \(13.73~\mathrm{Pa}\), respectively. This marked improvement indicates that nonlinear combinations of the geometric parameters account for a significant fraction of the response variability. The result further suggests that the governing transport mechanisms arise from coupled interactions among the design variables rather than from their independent contributions.
Among the nonlinear models, XGBoost and GPR both achieve excellent agreement with the CFD simulations on the internal test set. XGBoost attains \(R^2=0.9960\) for \(MI\) and \(R^2=0.9976\) for \(\Delta p\), whereas GPR further improves these values to \(R^2=0.9968\) and \(R^2=0.9999\), respectively. GPR also yields the smallest RMSE ($2.22\times10^{-3}$ for $MI$ and 0.54 Pa for $\Delta p$) and MAE ($8.5\times10^{-4}$ for $MI$ and 0.29 Pa for $\Delta p$), together with the lowest maximum error for $\Delta p$ (1.82 Pa) among all candidate models. The negligible prediction bias further indicates that the GPR surrogate reproduces the CFD responses without systematic over-prediction or under-prediction. 

The superior performance of GPR reflects the characteristics of the present design problem. The CFD-generated response surfaces remain continuous and smoothly varying throughout the investigated parameter space despite their strong nonlinearity. The Mat\'ern covariance kernel therefore interpolates efficiently between neighboring geometries while preserving localized variations associated with contraction-induced deformation and shear-thinning rheology. In contrast, XGBoost approximates the response using piecewise tree partitions. Although this strategy accurately captures the dominant nonlinear trends, it is inherently less effective for smoothly varying deterministic response surfaces. Figure~\ref{fig:surrogate_comparison}(b) further evaluates the models using the independent external holdout dataset, providing a more stringent assessment of predictive capability. GPR maintains excellent predictive accuracy, achieving \(R^2=0.9955\) for \(MI\) and \(R^2=0.9948\) for \(\Delta p\). PRR also generalizes well, with corresponding \(R^2\) values of 0.9920 and 0.9897. In contrast, XGBoost exhibits a noticeable reduction in predictive accuracy (\(R^2=0.7840\) for \(MI\) and \(R^2=0.6601\) for \(\Delta p\)), indicating weaker interpolation capability outside the training configurations. Linear regression remains the least accurate model, confirming that the geometric--transport relationship is intrinsically nonlinear.

The error distributions shown in Figs.~\ref{fig:surrogate_comparison}(c) and (d) provide additional insight into model robustness. GPR and PRR produce narrow error distributions for both objectives, indicating consistently accurate predictions throughout the investigated design space. In contrast, Linear Regression and XGBoost exhibit broader distributions with longer tails, reflecting larger prediction errors for specific geometric configurations. The predictive intervals for the pressure drop are consistently wider than those for the mixing index because \(\Delta p\) is strongly influenced by localized velocity gradients and viscous dissipation in the contraction regions, where small geometric changes can produce substantial variations in the pressure field. By contrast, \(MI\) is an integral measure of scalar transport over the entire channel length, which averages the effects of local flow variations. Consequently, \(\Delta p\) exhibits greater sensitivity to geometric perturbations and is therefore more challenging to predict accurately.

Overall, the combined quantitative metrics and error distributions identify GPR as the most accurate and robust surrogate model for the present application. Its excellent interpolation capability, minimal prediction bias, and consistently low prediction errors establish it as the most suitable surrogate for subsequent design-space exploration and multi-objective optimization.

\subsubsection{Predictive Accuracy on Independent CFD Data}

\begin{figure}[t]
    \centering
\includegraphics[width=0.8\linewidth]{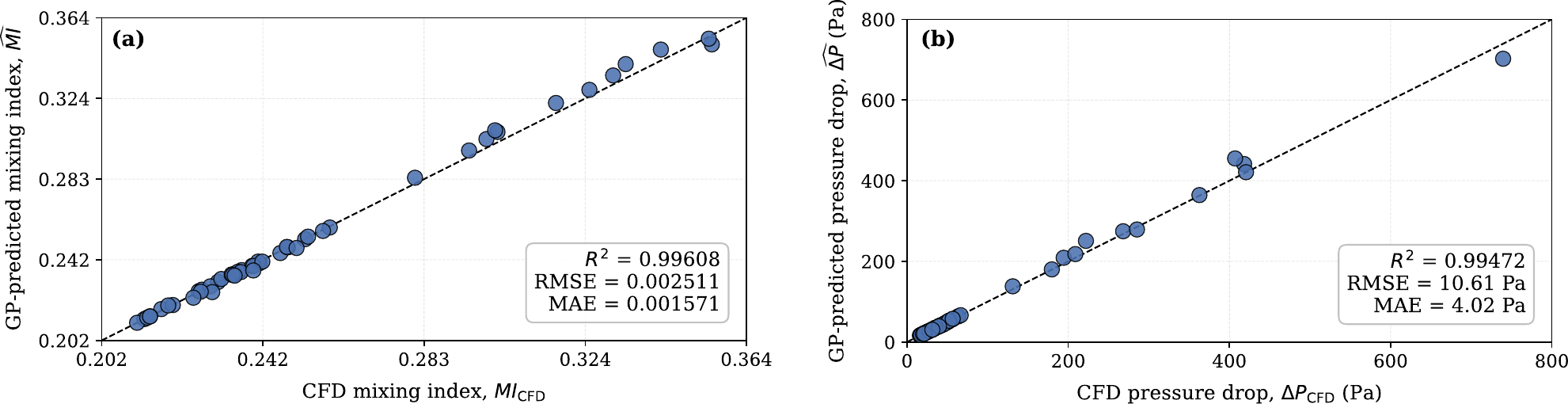}
    \caption{Parity plots comparing Gaussian Process Regression (GPR) predictions with CFD results for the external holdout dataset (\(N=49\)). (a) Mixing index (\(MI\)) and (b) pressure drop (\(\Delta p\)). The dashed line represents excellent agreement between surrogate predictions and CFD data. The high coefficients of determination (\(R^2\)), together with the low RMSE and MAE values, demonstrate the excellent predictive accuracy of the GPR surrogate model.}
    \label{fig:gpr_parity}
\end{figure}

Having identified GPR as the most accurate surrogate model, we next assess its predictive capability using the independent external holdout dataset. Figure~\ref{fig:gpr_parity} compares the GPR predictions with the corresponding CFD results for both the mixing index and pressure drop. Because none of the 49 holdout geometries participated in surrogate training, feature engineering, or hyperparameter selection, this dataset provides a stringent assessment of the model's ability to predict previously unseen configurations.

Figure~\ref{fig:gpr_parity} demonstrates excellent agreement between the GPR surrogate predictions and the CFD simulations for both the mixing index and pressure drop. The mixing-index surrogate achieves \(R^2=0.9961\), an RMSE of \(2.5\times10^{-3}\), and an MAE of \(1.6\times10^{-3}\), with predictions closely following the one-to-one parity line and exhibiting no noticeable bias. Similarly, the pressure-drop surrogate attains \(R^2=0.9947\), an RMSE of \(10.61~\mathrm{Pa}\), and an MAE of \(4.02~\mathrm{Pa}\). Although the pressure-drop predictions show slightly greater scatter, they remain tightly clustered about the parity line, reflecting the greater sensitivity of \(\Delta p\) to localized velocity gradients and viscous dissipation. Overall, the parity plots confirm that the GPR surrogate accurately captures the nonlinear relationship between the geometric design variables and both performance metrics. The parity plots also demonstrate that the surrogate accurately reproduces both low- and high-performance designs without loss of accuracy near the boundaries of the response space. No systematic overprediction or underprediction is observed for either objective, indicating that the selected Mat\'ern covariance function successfully captures the smooth yet highly nonlinear response surfaces generated by the CFD simulations. This agreement confirms that the GPR surrogate reconstructs the underlying geometric--transport relationship rather than simply interpolating individual training points.

To assess model robustness, we performed five-fold cross-validation using the selected hyperparameters. GPR consistently achieved \(R^2=0.997\pm0.002\) and \(\mathrm{RMSE}=0.00175\pm0.00075\) for \(MI\), and \(R^2=0.999\pm0.002\) with \(\mathrm{RMSE}=3.38\pm3.51~\mathrm{Pa}\) for \(\Delta p\), demonstrating stable performance across different training partitions. On the 49-case external holdout set, the empirical coverage of the nominal \(95\%\) predictive intervals reached \(91.8\%\) (45/49) for \(MI\) and \(93.9\%\) (46/49) for \(\Delta p\), indicating well-calibrated uncertainty estimates. The close agreement among the internal test, cross-validation, and independent external-validation results confirms GPR as a robust and unbiased surrogate for the present micromixer design problem. We note that Fig.~\ref{fig:surrogate_comparison} reports same-fit metrics, with all candidate models trained on the common 119-case training set to enable a fair comparison, whereas Fig.~\ref{fig:gpr_parity} presents the performance of the final GPR model refitted on all 169 primary cases and used in the subsequent optimization. Having established its predictive accuracy, we next examine how the interpolation errors are distributed throughout the multidimensional design space.

\subsubsection{Spatial Distribution of Prediction Errors}

\begin{figure}[t]
    \centering
    \includegraphics[width=0.8\linewidth]{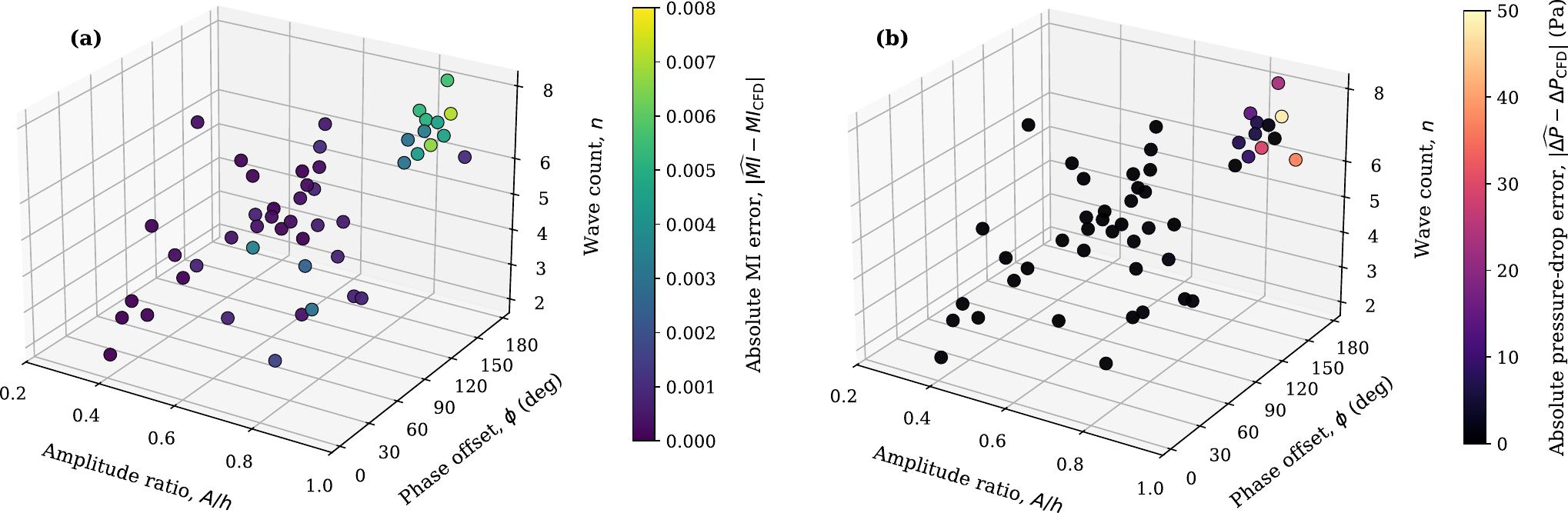}
    \caption{Error topology of the Gaussian Process Regression (GPR) surrogate model evaluated on the external holdout dataset. The prediction errors are visualized in the design space defined by the channel amplitude ratio (\(A/h\)), phase offset (\(\phi\)), and wave count (\(n\)). (a) Absolute mixing-index error, \(|\hat{MI}-MI_{\mathrm{CFD}}|\), and (b) absolute pressure-drop error, \(|\hat{\Delta p}-\Delta p_{\mathrm{CFD}}|\). The color scale represents the magnitude of the prediction error.}
    \label{fig:gpr_error_topology}
\end{figure}

While the parity plots establish the overall predictive accuracy of the GPR surrogate, they do not reveal how the prediction errors vary throughout the multidimensional design space. Figure~\ref{fig:gpr_error_topology} therefore maps the absolute prediction errors over the external holdout dataset as functions of the wall-amplitude ratio (\(A/h\)), phase offset (\(\phi\)), and wave count (\(n\)), providing a direct assessment of the surrogate's interpolation capability across previously unseen geometries.

Figure~\ref{fig:gpr_error_topology}(a) shows that the absolute errors in the predicted mixing index remain uniformly small throughout most of the design space, with the majority of the holdout cases exhibiting errors below \(5\times10^{-3}\). The error distribution does not exhibit large contiguous regions of poor performance, indicating that the surrogate accurately reconstructs the continuous dependence of mixing on the geometric design variables. Slightly larger deviations appear only for configurations combining relatively large wall amplitudes, high phase offsets, and shorter geometric wavelengths, where repeated contraction--expansion cycles generate stronger interfacial deformation and consequently a more rapidly varying response surface. A similar trend is observed for the pressure-drop prediction in Fig.~\ref{fig:gpr_error_topology}(b). The surrogate maintains low prediction errors over most of the parameter space, while the largest deviations remain confined to a small number of geometries located near the boundaries of the design domain. These configurations correspond to the strongest geometric modulation, where severe flow acceleration through the contraction regions produces localized peaks in velocity gradient and viscous dissipation. Because the pressure drop depends directly on these highly localized flow features, its response surface exhibits steeper gradients than that of the outlet mixing index, making accurate interpolation inherently more challenging.

Despite these localized increases, the prediction errors remain small relative to the overall ranges of both objective functions, with no systematic high-error regions across the design space. The consistently low errors for the external holdout dataset, including the excluded intermediate \(n=7\) geometries, demonstrate that the surrogate accurately interpolates between neighboring designs and captures the underlying response surfaces. Together with the parity plots, the error-topology analysis confirms that the GPR surrogate reliably predicts both \(MI\) and \(\Delta p\) across the design space, providing an accurate and computationally efficient alternative to CFD for subsequent response-surface analysis and multi-objective optimization.

\subsection{Design-Space Exploration}

\subsubsection{Response Surfaces of Mixing and Hydraulic Resistance}

\begin{figure}[t]
    \centering
\includegraphics[width=1\linewidth]{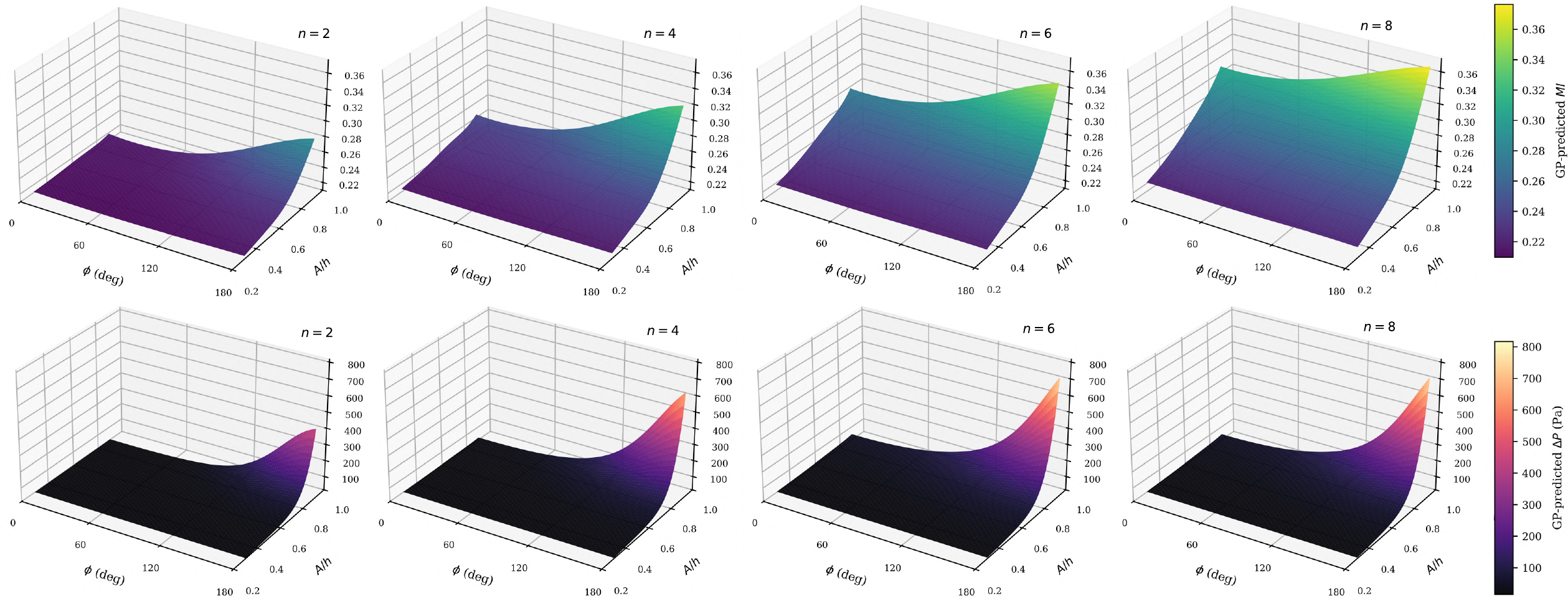}
    \caption{Gaussian Process Regression (GPR) response surfaces for the predicted mixing index (\(MI\), top row) and pressure drop (\(\Delta p\), bottom row) as functions of the amplitude ratio (\(A/h\)) and phase offset (\(\phi\)) for different wave counts (\(n=2,4,6,\) and \(8\)). The surfaces illustrate the combined influence of geometric parameters on micromixing performance and hydraulic resistance.}
    \label{fig:gpr_surfaces}
\end{figure}

To gain insight into the relationship between the geometric design parameters and the micromixer performance, we used the trained Gaussian Process Regression (GPR) surrogate to reconstruct continuous response surfaces over the design space. Figure~\ref{fig:gpr_surfaces} shows the predicted mixing index (\(MI\)) and pressure drop (\(\Delta p\)) as functions of the wall-amplitude ratio (\(A/h\)) and phase offset (\(\phi\)) for representative wave counts (\(n=2,\,4,\,6,\) and \(8\)). These response surfaces provide a comprehensive visualization of the multidimensional design space and reveal how geometric modulation simultaneously governs solute mixing and pressure drop.

The upper row of Fig.~\ref{fig:gpr_surfaces} shows that the mixing index generally increases with wall-amplitude ratio for all wave counts. Larger values of \(A/h\) strengthen the converging--diverging sections, producing greater interface stretching and increasing the interfacial area available for molecular diffusion. At a fixed wall-amplitude ratio, the phase offset controls the relative alignment of the upper and lower channel walls. As the phase offset increases, the contractions and expansions become progressively staggered, promoting stronger transverse fluid motion and repeated interface deformation that enhance scalar transport. Although this dependence is not strictly monotonic throughout the design space, larger phase offsets generally yield higher mixing indices. Increasing the wave count further improves mixing by shortening the geometric wavelength and increasing the number of contraction--expansion cycles within the fixed sinusoidal section, allowing repeated stretching of the concentration field. Consequently, the highest mixing indices are obtained for geometries combining large wall amplitudes, large phase offsets, and high wave counts. The lower row of Fig.~\ref{fig:gpr_surfaces} shows that the pressure drop exhibits similar trends but a stronger nonlinear dependence on the geometric parameters. Increasing the wall-amplitude ratio markedly increases hydraulic resistance by generating larger velocity gradients and greater viscous dissipation in the throat regions. The influence of the phase offset is comparatively weaker because it primarily redistributes the contraction--expansion sequence without substantially altering the minimum hydraulic cross-section. Increasing the wave count also raises the pressure drop by increasing the number of contraction--expansion events, leading to greater cumulative viscous losses along the channel.

Comparing the two response surfaces reveals the intrinsic trade-off governing the present design space. The geometric modifications responsible for increasing interfacial stretching also intensify the local velocity gradients that generate viscous dissipation. Consequently, the configurations that maximize mixing invariably produce larger pressure drops. This competition between transport enhancement and hydraulic resistance defines the Pareto-optimal design space explored in the subsequent optimization section. The smooth variation of both response surfaces further demonstrates that the selected GPR surrogate faithfully reconstructs the continuous dependence of the transport characteristics on the geometric parameters. The absence of artificial discontinuities confirms that the surrogate provides a physically consistent representation of the CFD response surfaces, enabling rapid exploration of the multidimensional design space without additional numerical simulations.

\subsubsection{Physical Interpretation of Feature Importance}

\begin{figure}[t]
    \centering
\includegraphics[width=0.9\linewidth]{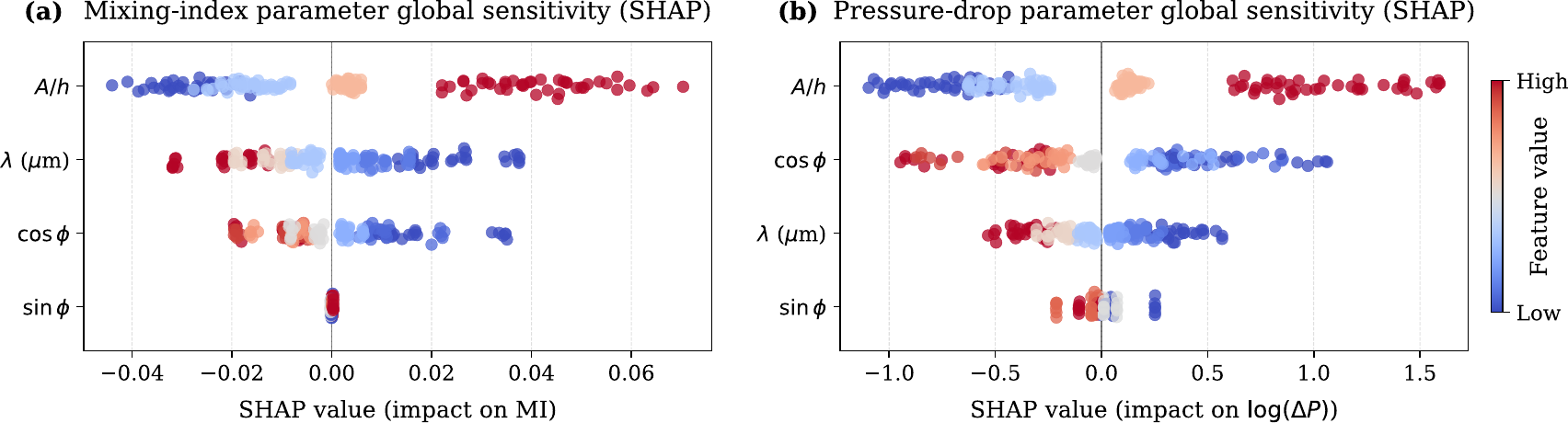}
    \caption{SHapley Additive exPlanations (SHAP) summary plots derived from the final refit XGBoost surrogate (a) Feature contributions to the predicted mixing index (\(MI\)) and (b) feature contributions to the predicted pressure drop ($\log(\Delta p)$). Each point represents a sample from the dataset, with the horizontal position indicating the SHAP value and the color denoting the corresponding feature magnitude.}    \label{fig:shap_summary}
\end{figure}

While the response surfaces identify the global trends within the design space, they do not quantify the relative contribution of each geometric parameter. Figure~\ref{fig:shap_summary} therefore presents SHAP values computed from the final refit XGBoost surrogate, providing a quantitative interpretation of the influence of each feature on the predicted mixing index and pressure drop.

For the mixing index, shown in Fig.~\ref{fig:shap_summary}(a), the grouped mean absolute SHAP attribution identifies the wall-amplitude ratio as the dominant design variable (50.0\%), followed by the geometric wavelength (28.3\%) and the joint phase encoding \((\sin\phi,\cos\phi)\) (21.6\%). Larger amplitude ratios consistently produce positive SHAP values, confirming that stronger wall modulation enhances mixing by increasing interface stretching and scalar homogenization, consistent with the CFD flow fields. Shorter wavelengths further improve mixing by introducing additional contraction--expansion cycles, whereas the phase offset exerts a weaker but generally positive influence through increased geometric asymmetry. The pressure-drop SHAP values shown in Fig.~\ref{fig:shap_summary}(b) indicates that the grouped mean absolute SHAP attribution ranks the wall-amplitude ratio as the most influential parameter (48.9\%), followed by the joint phase encoding \((\sin\phi,\cos\phi)\) (34.4\%) and the geometric wavelength (16.6\%). The dominant role of \(A/h\) reflects its direct control over the minimum hydraulic cross-section, which governs the contraction-induced velocity gradients and the associated viscous dissipation. Unlike the mixing index, the secondary importance shifts from wavelength to phase offset. The SHAP rankings are consistent with the trends identified from the CFD simulations and the surrogate response surfaces, where the wall-amplitude ratio dominates both mixing enhancement and pressure loss, while the phase offset and wavelength play secondary roles. This agreement indicates that the surrogate has learned physically meaningful relationships between the design variables and the target responses. However, SHAP explains how the surrogate attributes its predictions to the input features and should not be interpreted as establishing physical causality.

\subsection{Pareto Front and Optimal Design}
\begin{figure}[t]
    \centering
    \includegraphics[width=1\linewidth]{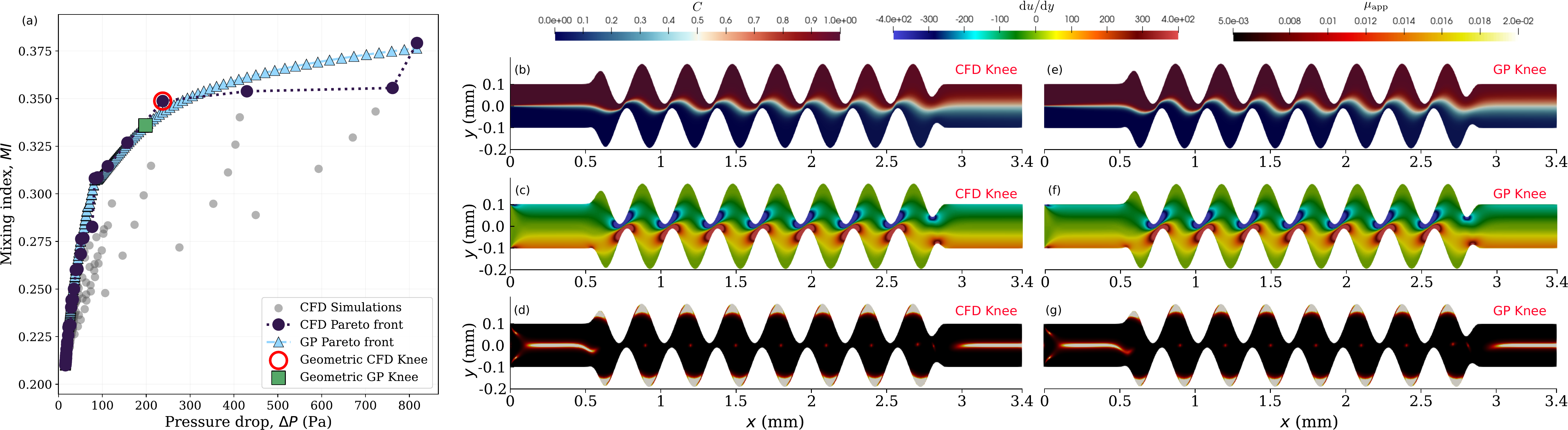}
    \caption{Comparison of CFD- and GP-derived Pareto fronts and corresponding knee-point designs. The left panel shows the Pareto-optimal trade-off between mixing index ($MI$) and pressure drop ($\Delta p$), with the CFD and GP knee solutions highlighted. The right panels compare the concentration field, velocity gradient, and apparent viscosity contours for the CFD-based (b--d) and GP-based (e--g) knee designs.}    
    \label{fig:pareto_front}
\end{figure}

After establishing the predictive accuracy of the Gaussian Process (GP) surrogate, we coupled it with the NSGA-II algorithm to identify micromixer geometries that simultaneously maximize the mixing index and minimize the pressure drop. Figure~\ref{fig:pareto_front} compares the Pareto fronts obtained directly from the CFD database and from the GP-assisted NSGA-II optimization. The close agreement between the two fronts demonstrates that the surrogate accurately reconstructs the trade-off between mixing enhancement and hydraulic resistance across the design space. The GP Pareto front closely follows the CFD Pareto boundary over the entire range of objective values, confirming that the surrogate preserves both the shape and topology of the underlying objective landscape.

The CFD Pareto front spans approximately \(0.21 \lesssim MI \lesssim 0.38\) and \(15.68 \lesssim \Delta p \lesssim 817.27~\mathrm{Pa}\). As shown in Fig.~\ref{fig:pareto_front}(a), the Pareto front exhibits a pronounced concave shape. At low pressure drops, relatively small increases in \(\Delta p\) produce substantial improvements in the mixing index because moderate geometric modulation significantly increases interfacial stretching while introducing only a modest hydraulic resistance. Beyond this region, however, the slope of the Pareto front decreases markedly, indicating diminishing returns in mixing performance. Additional increases in wall modulation continue to intensify contraction-induced deformation, but the corresponding gain in mixing becomes progressively smaller, whereas the pressure drop increases rapidly owing to the accumulation of viscous dissipation within successive contraction--expansion units. The knee point therefore represents the most favorable compromise between
transport enhancement and hydraulic resistance, marking the onset of
diminishing returns beyond which further mixing gains demand disproportionately
larger increases in pumping power.

For the GP-selected knee geometry (\(A/h = 0.90\), \(\phi = 106.77^{\circ}\), \(n = 8\)), the GP surrogate predicts \(MI = 0.335\) and \(\Delta p = 194.9~\mathrm{Pa}\). Independent CFD simulations confirm these predictions with \(MI = 0.338\) and \(\Delta p = 195.4~\mathrm{Pa}\), corresponding to relative errors of only \(0.89\%\) and \(0.26\%\), respectively. The nominal \(95\%\) GP posterior predictive intervals encompass both CFD values, demonstrating that the surrogate reliably identifies optimal designs. Compared with the CFD knee solution (\(A/h = 0.90\), \(\phi = 120.00^{\circ}\), \(n = 8\); \(MI = 0.349\), \(\Delta p = 237.5~\mathrm{Pa}\)), the GP-selected design reduces the pressure drop by \(17.9\%\) while decreasing the mixing index by only \(3.2\%\). Although the two knee solutions differ only in phase offset, they occupy the same high-amplitude, high-wave-count region of the Pareto front, and the GP-selected design provides the more hydraulically efficient trade-off. Figure~\ref{fig:pareto_front}(b--g) compares the concentration, velocity gradient, and apparent-viscosity fields of the GP-selected and corresponding CFD-database designs. The two geometries exhibit closely similar concentration evolution, velocity gradients, and shear-thinning behavior, indicating that the surrogate reproduces the same transport mechanisms, hydraulic response, and rheological characteristics captured by the high-fidelity CFD simulations. This close agreement confirms that the surrogate identifies a physically consistent optimum rather than merely fitting the training data. Consequently, the GP-assisted framework enables rapid and reliable optimization of shear-thinning micromixers while reducing the computational cost by several orders of magnitude compared with repeated high-fidelity CFD simulations.

We further validated the GP surrogate by independently simulating the pressure-constrained optimal designs listed in Table~\ref{tab:pressure_constrained_cfd_confirmation}. The GP surrogate predicted the pressure drop with relative errors below \(1\%\) for all five designs, corresponding to a mean relative absolute error of approximately \(0.38\%\), while the mixing-index predictions differed from the CFD values by at most \(2.74\%\). These results demonstrate that the surrogate accurately identifies designs that satisfy the prescribed pressure constraints while reliably predicting the corresponding mixing performance across the constrained design space.
\begin{table*}[t]
\centering
\caption{Independent CFD confirmation of the pressure-constrained designs}
\label{tab:pressure_constrained_cfd_confirmation}
\setlength{\tabcolsep}{4.5pt}
\begin{tabular}{cccccccccc}
\toprule
$\Delta p_{\mathrm{lim}}$
& $A/h$
& $\phi$
& $n$
& $MI_{\mathrm{GP}}$
& $MI_{\mathrm{CFD}}$
& $e_{MI}$
& $\Delta p_{\mathrm{GP}}$
& $\Delta p_{\mathrm{CFD}}$
& $e_{\Delta p}$ \\
(Pa)
& {}
& ($^\circ$)
& {}
& {}
& {}
& (\%)
& (Pa)
& (Pa)
& (\%) \\
\midrule
100 & 0.9 &  46.05 & 8 & 0.310585 & 0.311075 & 0.157 &  99.183 &  99.145 & 0.039 \\
150 & 0.9 &  84.92 & 8 & 0.323961 & 0.324135 & 0.054 & 147.670 & 147.638 & 0.022 \\
200 & 0.9 & 107.89 & 8 & 0.335858 & 0.338586 & 0.806 & 197.919 & 198.318 & 0.201 \\
300 & 0.9 & 132.12 & 8 & 0.350289 & 0.360148 & 2.737 & 294.459 & 292.532 & 0.659 \\
400 & 0.9 & 145.89 & 8 & 0.358778 & 0.367018 & 2.245 & 392.158 & 388.358 & 0.979 \\
\bottomrule
\end{tabular}
\end{table*}
The close correspondence between the CFD and GP Pareto fronts also demonstrates the interpolation capability of the surrogate. Rather than reproducing only the discrete CFD database, the GP model reconstructs the continuous objective landscape and enables efficient exploration of previously unsampled regions of the design space. Consequently, surrogate-assisted optimization substantially reduces the computational cost associated with exhaustive CFD searches while preserving the physical trade-offs governing micromixer performance.

\section{Conclusions}
\label{sec:conclusions}
We developed a surrogate-assisted computational framework for the rapid design and multi-objective optimization of sinusoidal converging--diverging micromixers transporting shear-thinning Carreau--Yasuda fluids. We generated a CFD database comprising 218 simulations to quantify the effects of the wall-amplitude ratio (\(A/h\)), phase offset (\(\phi\)), and wave count (\(n\)) on the mixing index (\(MI\)) and pressure drop (\(\Delta p\)). Among the four surrogate models investigated, Gaussian Process Regression (GPR) delivered the highest predictive accuracy and captured the nonlinear relationships between the geometric parameters and the transport characteristics.

Our CFD simulations showed that sinusoidal geometric modulation enhances mixing through the combined effects of interface stretching, elevated shear rates, and shear-thinning rheology. Successive contraction--expansion cycles continuously stretched and elongated the concentration interface, increasing the interfacial area available for molecular diffusion. At the same time, elevated shear rates reduced the apparent viscosity through the Carreau--Yasuda constitutive model and promoted further interface deformation. These coupled mechanisms enhanced scalar transport while increasing hydraulic resistance. We found that the wall-amplitude ratio dominated both mixing enhancement and pressure loss, whereas the phase offset and wave count provided secondary control by altering the spatial distribution and frequency of fluid deformation.

We used the GPR surrogate to reconstruct the CFD response surfaces and efficiently explore the multidimensional design space. By coupling the surrogate with NSGA-II, we reproduced the CFD Pareto front and identified optimal geometries that balanced mixing performance and pressure loss. Independent CFD simulations confirmed the surrogate predictions of \(MI\) and \(\Delta p\) with excellent agreement. Although the surrogate predicts only these scalar objectives, the confirmation simulations reproduced the concentration, pressure, velocity gradient, and apparent-viscosity fields of the corresponding CFD-database designs, demonstrating that the surrogate identifies physically meaningful optimal geometries that preserve the underlying transport mechanisms rather than merely fitting the objective values.

Overall, the proposed CFD--machine-learning framework offers a fast, accurate, and physically consistent strategy for designing shear-thinning micromixers while substantially reducing the computational cost of repeated high-fidelity CFD simulations. Although we focused on two-dimensional Carreau--Yasuda flows and interpolation within the sampled design space, we can readily extend the framework to more complex microfluidic systems. In future work, we will apply the framework to three-dimensional micromixer geometries, incorporate additional generalized-Newtonian and viscoelastic constitutive models, and validate the optimized designs experimentally to broaden the applicability of surrogate-assisted microfluidic design.

\section{Acknowledgments}
B.M. acknowledges support from the Department of Science and Technology(DST), India, through the INSPIRE Faculty Fellowship (Grant No. DST/INSPIRE/04/2024/003069).

\section{Data availability statement}

The datasets generated and analyzed during the current study, together with the machine-learning surrogate models and NSGA-II optimization codes are publicly available in the \texttt{surrogate-micromixer-optimization} GitHub repository:
\url{https://github.com/BimalenduMahapatra/surrogate-micromixer-optimization}.\\
    
\hrule
\bibliographystyle{unsrtnat}
\bibliography{references}

\end{document}